\documentclass[aps,twocolumn,showpacs]{revtex4}
\usepackage{epsfig}
\usepackage{amsbsy}
\usepackage{amsmath}
\usepackage{amsfonts}
\usepackage{graphicx}
\usepackage{latexsym}

\begin{document}

\title{Analytical results for entanglement in the five-qubit anisotropic Heisenberg model}
\author{XiaoGuang Wang}
\affiliation{Zhejiang Institute of Modern Physics, Department of
Physics, Zhejiang University, HangZhou 310027, China}
\date{\today}
\begin{abstract}
We solve the eigenvalue problem of the five-qubit anisotropic
Heisenberg model, without use of Bethe's Ansatz, and give
analytical results for entanglement and mixedness of two
nearest-neighbor qubits. The entanglement takes its maximum at
$\Delta=1$ ($\Delta>1$) for the case of zero (finite) temperature
with $\Delta$ being the anisotropic parameter. In contrast, the
mixedness takes its minimum at $\Delta=1$ ($\Delta>1$) for the
case of zero (finite) temperature.
\end{abstract}
\pacs{03.65.Ud, 03.67.-a}
\maketitle

Recently, the study of entanglement properties of many-body
systems has received much attention~\cite{M_Nielsen}-\cite{QPT_GVidal}. To obtain analytical results for entanglement, one may consider
the case of infinite lattice or a small lattice with a few qubits.
It is hard to get some analytical results between these two
extreme cases.

It was well-known that the anisotropic Heisenberg model can be
solved formally by Bethe's Ansatz method~\cite{Bethe,Yang} for
arbitrary number of qubits $N$, however, we have to solve a set of
transcendental equations. For $N\le 7$, the isotropic Heisenberg
Hamiltonian can be analytically solved~\cite{Kouzoudis,Schnack}.
Here, we give the analytical results of the eigenvalues of the
anisotropic Heisenberg model with $N=5$, without use of Bethe's
Ansatz, from which the analytical expressions for entanglement and
mixedness of two nearest-neighbor qubits are readily obtained.

It is interesting to see that the entanglement properties of a
pair of nearest-neighbor qubits at a finite temperature is
completely determined by the partition function. The entanglement,
quantified by the concurrence~\cite{Conc}, relates to the
partition function $Z$ via~\cite{Glaser,Gu,WangPaolo}
\begin{equation}
C=\max\left(0,\Big|\frac{U}{N}-\frac{1}{2}-\frac{\Delta G_{zz}}2\Big|-\frac{G_{zz}}{2}-\frac{1}{2}\right)\label{c1}
\end{equation}
with
\begin{equation}
U=-\frac{\partial \ln Z}{\partial\beta}\label{c2}
\end{equation}
being the internal energy, and
\begin{equation}
G_{zz}=\text{Tr}[\exp(-\beta H)\sigma_{iz}\sigma_{i+1z}]/Z=-\frac{2}{N\beta}\frac{\partial\ln Z}{\partial\Delta}\label{c3}
\end{equation}
being the correlation function. Here, $\beta=1/T$ and the
Boltzmann's constant $k=1$. Thus, once we know the eigenenergies
versus the temperature and the anisotropic parameter, we can
completely determine the entanglement.

There exists another concept, the mixedness of a state, is central
in quantum information theory~\cite{Jaeger}. For instance, Bose
and Vedral have shown that entangled states become useless for
quantum teleportation on exceeding a certain degree of
mixedness~\cite{Bose}. Mixedness is also related to quantum
entanglement. We will study both the entanglement and mixedness
properties.

{\em Eigenvalue problem.} The anisotropic Heisenberg Hamiltonian
is given by
\begin{align}
H=&\frac{J}2\sum_{i=1}^{N}\left(1+\sigma_{ix}\sigma_{i+1x}+\sigma_{iy}\sigma_{i+1y}+\Delta\sigma_{iz}\sigma_{i+1z}\right)\nonumber\\
=&J\sum_{i=1}^{N}\left(S_{i,i+1}+\frac{\Delta-1}{2}\sigma_{iz}\sigma_{i+1z}\right),
\label{xxz}
\end{align}
where $S_{j,j+1}=${}$\frac 12\left( 1+\vec{\sigma}_i\cdot \vec{\sigma}%
_{i+1}\right) $ is the swap operator between qubit $i$ and $j$, $\vec{\sigma}%
_i=(\sigma _{ix},\sigma _{iy},\sigma _{iz})$ is the vector of
Pauli matrices, and $J$ is the exchange constant. We have assumed
the periodic boundary condition, i.e., $N+1\equiv 1$. In the
following discussions, we also assume $J=1$ (antiferromagnetic
case) and $\Delta\ge 0$.

Since we impose the periodic boundary condition, the Hamiltonian
is translational invariant, i.e., $[H,T]=0$, where $T$ is the
cyclic right shift operator defined as
\begin{equation}
T|m_1,\cdots,m_{N-1},m_N\rangle=|m_N,m_1,\cdots,m_{N-1}\rangle.
\end{equation}
The translational invariant symmetry can be used to reduce the
Hamiltonian matrix to smaller submatrices by a factor of
$N$~\cite{Lin}.

Now we focus our attention to five-qubit settings, and solve the
eigenvalue problem of the anisotropic Heisenberg model.  Since
$[H,J_z]=0$, the whole 32-dimentional Hilbert space can be divided
into invariant subspaces spanned by vectors with a fixed number of
reversed spins. Then, the largest subspace is 10-dimensional with
2 or 3 reversed spins. Here, $J_z=\sum_{i=1}^5\sigma_{iz}/2$. Due
to the symmetry $[H,\Sigma_x]=0$, it is sufficient to solve the
eigenvalue problems in the subspaces with $r$ reversed spins,
where $r\in\{0,1,2\}$ and $
\Sigma_x=\sigma_x\otimes\sigma_x\otimes\sigma_x\otimes\sigma_x\otimes\sigma_x.
$ By using the translational invariance, we can further reduce the
Hamiltonian matrix to $2\times 2$ submatrices, and the eigenvalue
problem can be readily solved.

The subspace with $r=0$ only contains one vector $|00000\rangle$,
which is the eigenvector with eigenvalue
\begin{equation}
E_0=5[1+(\Delta-1)/2].\label{e1}
\end{equation}
The subspace with $r=1$ is spanned by five basis vectors
$\{T^n|10000\rangle, n\in \{0,1,2,3,4\}\}$. Considering the
translational invariance of the Hamiltonian, we choose another
basis given by
\begin{equation}
|\varphi_k\rangle=\sum_{n=0}^4 \omega^n_kT^n |10000\rangle,
\end{equation}
where $\omega_k=e^{\frac{i2k\pi}{5}}, k\in\{1,2,3,4,5\}$. It can
be checked that states $|\varphi_k\rangle$ are eigenstates of $T$
with eigenvalues $\omega_k^{-1}$, and are also eigenstates of
Hamiltonian $H$ with eigenvalues given by
\begin{equation}
E_{1,k}=3+\frac{\Delta-1}{2}+2\cos(2k\pi/5).\label{e2}
\end{equation}

For the case of $r=2$, we choose the following basis for
10-dimensional subspace
\begin{align}
|\psi_k\rangle=\sum_{n=0}^4 \omega_k^nT^n |11000\rangle,\nonumber\\
|\phi_k\rangle=\sum_{n=0}^4 \omega_k^nT^n |10100\rangle.
\end{align}
States $|\psi_k\rangle$ and $|\phi_k\rangle$ span an invariant
$2\times 2$ subspace under the action of Hamiltonian $H$. In this
subspace, the Hamiltonian can be written as
\begin{equation}
H_k=\left(
\begin{array}{ll}
3+\frac{\Delta-1}{2}&1+\omega_k^{-1}\\
1+\omega_k&1-\frac{3(\Delta-1)}{2}+\omega_k^2+\omega_k^{-2}
\end{array}
\right).
\end{equation}
Then, from the above equation, the eigenvalues are obtained as
\begin{align}
E_{2,k\pm}=&\frac{5-\Delta+2\cos(4k\pi/5)}{2}\nonumber\\
\pm & \sqrt{[\Delta-\cos(4k\pi/5)]^2+2[1+\cos(2k\pi/5)]}.
\label{e3}
\end{align}
Thus, all eigenvalues are obtained for the five-spin anisotropic
Heisenberg models. We see that eigenstates are at least two-fold
degenerate due to the symmetry $[H,\Sigma_x]=0$. Although the
eigenstates can be easily obtained, they are not given explicitly
here as the knowledge of eigenvalues is sufficient for discussions
of entanglement and mixedness properties.

\begin{figure}
\includegraphics[width=0.45\textwidth]{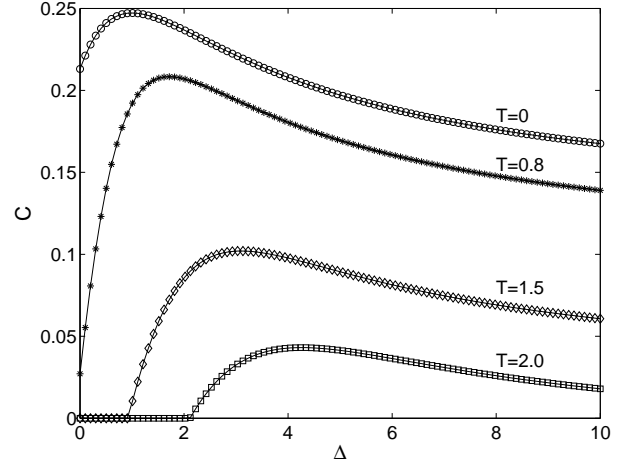}
\caption{The concurrence versus $\Delta$ for different temperatures.}
\end{figure}

{\em Entanglement and mixedness.} From
Eqs.~(\ref{c1})--(\ref{c3}), the concurrence is determined by the
partition function and its derivative with respect to the
anisotropic parameter $\Delta$. As we have obtained all
eigenvalues of five-spin Hamiltonian $H$, from
Eqs.~(\ref{e1})-(\ref{e2}), it follows that
\begin{equation}
Z=2\left[e^{-\beta E_0}+\sum_{k=1}^5 \left( e^{-\beta E_{1,k}}+
e^{-\beta E_{2,k+}}+ e^{-\beta E_{2,k-}}
\right)\right].\label{zzz}
\end{equation}
Substituting the above equation into Eqs.~(\ref{c2}) and (\ref{c3}) yields
\begin{align}
U=&\frac{2}{Z}\Big[E_0e^{-\beta E_0}+\sum_{k=1}^5 (
E_{1,k}e^{-\beta E_{1,k}}
\nonumber\\
&+E_{2,k+}e^{-\beta E_{2,k+}}+ E_{2,k-}e^{-\beta E_{2,k-}} )\big],
\end{align}
and
\begin{align}
G_{zz}=&\frac{2}{5Z}
\Big[\frac{5}{2}e^{-\beta E_0}+\sum_{k=1}^5 \big(
\frac{1}{2}e^{-\beta E_{1,k}} \nonumber\\
+&\frac{\partial E_{2,k+}}{\partial \Delta} e^{-\beta E_{2,k+}}+
\frac{\partial E_{2,k-}}{\partial \Delta}e^{-\beta E_{2,k-}}
\big)\Big].
\end{align}
with
\begin{equation}
\frac{\partial E_{2,k\pm}}{\partial \Delta}=-\frac{1}{2}\pm
\frac{\Delta-\cos(4k\pi/5)} {\sqrt{[ \Delta-\cos(4k\pi/5)
]^2+2[1+\cos(2k\pi/5)]}}.\label{par}
\end{equation}
Then, the analytical expressions of the internal engery $U$ and
the correlation function $G_{zz}$ are obtained, and thus,
according to Eq.~(\ref{c1}), we get the analytical expression of
the concurrence for the thermal state.

At zero temperature, the system is in the ground state, and for
this case,  Eq.~(\ref{c1}) reduces to
\begin{equation}
C=\max\left(0,\big|\frac{E_{gs}}{N}-\frac{1}{2}-\frac{\Delta}{N}\frac{\partial
E_{gs}}{\partial \Delta}\big| -\frac{1}{N}\frac{\partial
E_{gs}}{\partial \Delta}-\frac{1}{2} \right), \label{cccc}
\end{equation}
where $E_{gs}$ denotes the ground-state energy and $N=5$. In the
derivation of the above equation, we have used the relation
\begin{equation}
G_{zz}=\frac{2}{N}\langle\frac{\partial H}{\partial\Delta}\rangle
=\frac{2}{N}\frac{\partial E_{gs}}{\partial\Delta},
\end{equation}
which is valid for the ground state. For our five-qubit model,
$E_{gs}=E_{2,1-}$ and the derivative of $E_{gs}$ with respect to
$\Delta$ is  given by Eq.~(\ref{par}). Thus, the analytical
expression for the entanglement of ground state is obtained. If $
\frac{E_{gs}}{N}-\frac{1}{2}-\frac{\Delta}{N}\frac{\partial
E_{gs}}{\partial \Delta}\le 0$, then Eq.~(\ref{cccc}) reduces to
\begin{equation}
C=-\frac{E_{gs}}{N}+\frac{\Delta-1}{N}\frac{\partial
E_{gs}}{\partial\Delta},
\end{equation}
where we have ignored the max function. Taking derivative with
respect to $\Delta$ on both sides of the above equation leads to
\begin{equation}
\frac{\partial C}{\partial
\Delta}=\frac{\Delta-1}{N}\frac{\partial^2E_{gs}}{\partial^2\Delta}.
\end{equation}
Then, it is direct to check from the above equation and the ground
state energy $E_{2,1-}$ (\ref{e3}) that the derivative is zero
when $\Delta=1$. Thus, the concurrence takes its extreme value at
the point of $\Delta=1$ for the ground state.

From the analytical results for the concurrence, we numerically
plot the concurrence versus the anisotropic parameter for
different temperatures in Fig.~1. We observe that the concurrence
takes its maximum when $\Delta=1$. This point correspond to the
critical point of metal-insulation transition~\cite{Gu}. However,
for finite temperature, the concurrence reaches its maximum when
$\Delta>1$. For finite temperatures (for instance, $T=2.0$), we
find a threshold value of the anisotropic parameter
$\Delta_\text{th}$, before which there is no pairwise
entanglement. The threshold value increase as temperature
increases.

\begin{figure}
\includegraphics[width=0.45\textwidth]{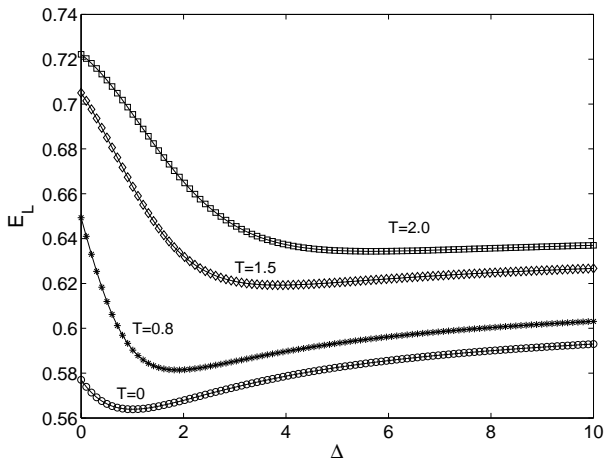}
\caption{The linear entropy versus $\Delta$ for different temperatures.}
\end{figure}

Next, we study mixedness properties of the thermal state and
ground state. The mixedness of a state $\varrho$ can be quantified
by the linear entropy given by
$
E_L=1-\text{Tr}(\varrho^2).
$
Then, for arbitrary number of qubits, the linear entropy of the
state of two nearest qubits is given by
\begin{equation}
E_L(T)=1-\frac{1}4 \left[
2\Big(\frac{U}{N}-\frac{1}{2}-\frac{\Delta
G_{zz}}2\Big)^2+{G_{zz}}^{2}+{1} \right]\label{ee1}
\end{equation}
for the thermal state, and
\begin{align}
E_L(T=0)=&1-\frac{1}4 \Big[
2\Big(\frac{E_{gs}}{N}-\frac{1}{2}-\frac{\Delta}N \frac{\partial
E_{gs}}{\partial\Delta}\Big)^2 \nonumber\\
&+\frac{4}{N^2}\Big(\frac{\partial
E_{gs}}{\partial\Delta}\Big)^{2}+{1} \Big]\label{eee1}
\end{align}
for the ground state, respectively. We see that the mixedness of
the thermal state is also completely determined by the partition
function, and the mixedness of the ground state is determined by
the ground-state energy and its first-order derivative with
respect to the anisotropic parameter. Thus, the analytical
expressions of the linear entropy are obtained.

\begin{figure}
\includegraphics[width=0.45\textwidth]{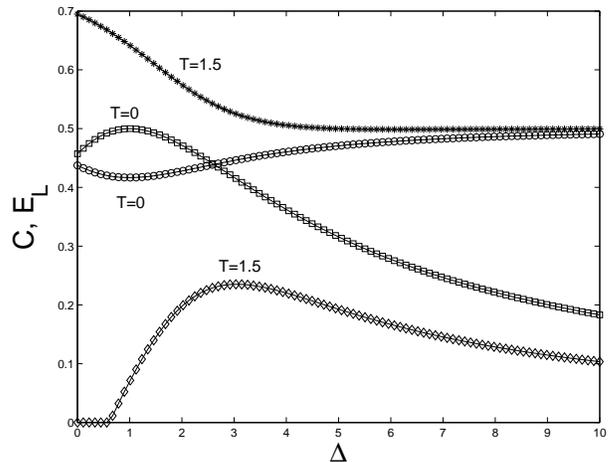}
\caption{The concurrence (square and diamond lines) and linear
entropy (circle and star lines) versus $\Delta$ for different
temperatures in the four-qubit model.}
\end{figure}

In Fig.~2, we numerically calculated the linear entropy versus
$\Delta$ for different temperatures. In contrast to the
entanglement, the mixedness of the ground state displays a minimum
when $\Delta=1$. For the case of finite temperatures, the
mixedness takes its minimum when $\Delta>1$. For lower
temperatures (for instance, $T\le 0.2$), numerical results show
that the maximum of the concurrence occurs nearly at the same
value of $\Delta$ as the minimum of the mixedness. It seems that
the more the pairwise entanglement, the less the mixedness.
However, for higher temperatures, the maximum of the concurrence
and the minimum of the mixedness do not occur at the same
$\Delta$. For instance, when $T=1.5$, the concurrence takes its
maximum at $\Delta=3.1037$, and the mixedness takes its minimum at
$\Delta=3.8525$. This signifies that it is not always true that
the more the pairwise entanglement and the less the mixedness.

For the case of four qubits, the anisotropic Heisenberg model is
also exactly solvable using the same method as above. Here, we
make a comparison of the four-qubit and five-qubit cases. The
exact ground-state energy and its derivative with respect to
$\Delta$ are given by
\begin{equation}
E_{gs}=2-\Delta-\sqrt{\Delta^2+8},\; \frac{\partial
E_{gs}}{\partial\Delta}=-1-\frac{\Delta}{\sqrt{\Delta^2+8}}.
\end{equation}
Substituting the above equation into Eqs.~(\ref{cccc}) and
(\ref{eee1}), we obtain the ground-state concurrence and linear
entropy as
\begin{align}
C(T=0)=&\frac{\Delta+8}{4\sqrt{\Delta^2+8}}-\frac{1}{4},\nonumber\\
E_L(T=0)=&\frac{11}{16}-\frac{\Delta}{8\sqrt{\Delta^2+8}}-\frac{\Delta^2+32}{16(\Delta^2+8)}.
\end{align}
From the above analytical expressions, it is straightforward to
check that the concurrence takes its maximum and the linear
entropy takes its minimum at the point of $\Delta=1$, which is
exactly the same feature as that in the five-qubit model. For
finite temperatures, in Fig.~3, we give numerical calculations of
the concurrence and the linear entropy. We see that they displays
similar behaviours as those in the five-qubit model. For instance,
for $T>0$, the maximum pairwise entanglement and the minimum
mixedness occur at $\Delta>1$.

{\em Conclusion.} In conclusion, we have obtained the analytical
results for the entanglement and mixedness  in the five-qubit
anisotropic Heisenberg model. The exact eigenspectrum is obtained,
and entanglement and mixedness properties can be completely
determined by the eigenvalues of the system, irrespective of the
eigenstates. The method adopted here can be applied to the
anisotropic Heisenberg model with more than five qubits (for
instance, 6 or 7 qubits). However, the analytical expressions for
eigenvalues, entanglement, and mixedness are expected to be more
complicated.

We have made numerical calculations, and show that the
entanglement takes its maximum at $\Delta=1$ ($\Delta>1$) for the
case of zero (finite) temperature. In contrast, the mixedness
takes its minimum at $\Delta=1$ ($\Delta>1$) for the case of zero
(finite) temperature. From our analysis, we conjecture that it is
a general feature that at zero temperature the entanglement takes
its maximum and the mixedness takes its minimum when $\Delta=1$
for any number of qubits. This conjecture is supported by our
analytical results for four and five qubits, and by numerical
results for the number of qubits being as large as 1280~\cite{Gu}.
The Heisenberg chains not only displays rich entanglement
features, but also have useful applications such as the quantum
state transfer~\cite{M_Sub}. Experimentally, it was found that
entanglement is crucial to describing magnetic behaviors in a
quantum spin system~\cite{Exp}. So, the study of entanglement and
mixedness properties in the Heisenberg models will strength our
understanding of other quantum features of magnetic systems.

\acknowledgements The author thanks Y. Q. Li, C. P. Sun, and Z.
Song for helpful discussions.

\end{document}